\documentstyle[twocolumn,prl,aps]{revtex}
\def\PsfigVersion{1.9}
\ifx\undefined\psfig\else \fi

\let\LaTeXAtSign=\@
\let\@=\relax
\edef\psfigRestoreAt{\catcode`\@=\number\catcode`@\relax}
\catcode`\@=11\relax
\newwrite\@unused
\def\ps@typeout#1{{\let\protect\string\immediate\write\@unused{#1}}}
\ps@typeout{psfig/tex \PsfigVersion}

\def\figurepath{./}

\def\@nnil{\@nil}
\def\@empty{}
\def\@psdonoop#1\@@#2#3{}
\def\@psdo#1:=#2\do#3{\edef\@psdotmp{#2}\ifx\@psdotmp\@empty \else
    \expandafter\@psdoloop#2,\@nil,\@nil\@@#1{#3}\fi}
\def\@psdoloop#1,#2,#3\@@#4#5{\def#4{#1}\ifx #4\@nnil \else
       #5\def#4{#2}\ifx #4\@nnil \else#5\@ipsdoloop #3\@@#4{#5}\fi\fi}
\def\@ipsdoloop#1,#2\@@#3#4{\def#3{#1}\ifx #3\@nnil 
       \let\@nextwhile=\@psdonoop \else
      #4\relax\let\@nextwhile=\@ipsdoloop\fi\@nextwhile#2\@@#3{#4}}
\def\@tpsdo#1:=#2\do#3{\xdef\@psdotmp{#2}\ifx\@psdotmp\@empty \else
    \@tpsdoloop#2\@nil\@nil\@@#1{#3}\fi}
\def\@tpsdoloop#1#2\@@#3#4{\def#3{#1}\ifx #3\@nnil 
       \let\@nextwhile=\@psdonoop \else
      #4\relax\let\@nextwhile=\@tpsdoloop\fi\@nextwhile#2\@@#3{#4}}
\ifx\undefined\fbox
\newdimen\fboxrule
\newdimen\fboxsep
\newdimen\ps@tempdima
\newbox\ps@tempboxa
\long\def\fbox#1{\leavevmode\setbox\ps@tempboxa\hbox{#1}\ps@tempdima\fboxrule
    \advance\ps@tempdima \fboxsep \advance\ps@tempdima \dp\ps@tempboxa
   \hbox{\lower \ps@tempdima\hbox
  {\vbox{\hrule height \fboxrule
          \hbox{\vrule width \fboxrule \hskip\fboxsep
          \vbox{\vskip\fboxsep \box\ps@tempboxa\vskip\fboxsep}\hskip 
                 \fboxsep\vrule width \fboxrule}
                 \hrule height \fboxrule}}}}
\fi
\newread\ps@stream
\newif\ifnot@eof       
\newif\if@noisy        
\newif\if@atend        
\newif\if@psfile       
{\catcode`\%=12\global\gdef\epsf@start{
\def\epsf@PS{PS}
\def\epsf@getbb#1{%
\openin\ps@stream=#1
\ifeof\ps@stream\ps@typeout{Error, File #1 not found}\else
   {\not@eoftrue \chardef\other=12
    \def\do##1{\catcode`##1=\other}\dospecials \catcode`\ =10
    \loop
       \if@psfile
	  \read\ps@stream to \epsf@fileline
       \else{
	  \obeyspaces
          \read\ps@stream to \epsf@tmp\global\let\epsf@fileline\epsf@tmp}
       \fi
       \ifeof\ps@stream\not@eoffalse\else
       \if@psfile\else
       \expandafter\epsf@test\epsf@fileline:. \\%
       \fi
          \expandafter\epsf@aux\epsf@fileline:. \\%
       \fi
   \ifnot@eof\repeat
   }\closein\ps@stream\fi}%
\long\def\epsf@test#1#2#3:#4\\{\def\epsf@testit{#1#2}
			\ifx\epsf@testit\epsf@start\else
\ps@typeout{Warning! File does not start with `\epsf@start'.  It may not be a PostScript file.}
			\fi
			\@psfiletrue} 
{\catcode`\%=12\global\let\epsf@percent=
\long\def\epsf@aux#1#2:#3\\{\ifx#1\epsf@percent
   \def\epsf@testit{#2}\ifx\epsf@testit\epsf@bblit
	\@atendfalse
        \epsf@atend #3 . \\%
	\if@atend	
	   \if@verbose{
		\ps@typeout{psfig: found `(atend)'; continuing search}
	   }\fi
        \else
        \epsf@grab #3 . . . \\%
        \not@eoffalse
        \global\no@bbfalse
        \fi
   \fi\fi}%
\def\epsf@grab #1 #2 #3 #4 #5\\{%
   \global\def\epsf@llx{#1}\ifx\epsf@llx\empty
      \epsf@grab #2 #3 #4 #5 .\\\else
   \global\def\epsf@lly{#2}%
   \global\def\epsf@urx{#3}\global\def\epsf@ury{#4}\fi}%
\def\epsf@atendlit{(atend)} 
\def\epsf@atend #1 #2 #3\\{%
   \def\epsf@tmp{#1}\ifx\epsf@tmp\empty
      \epsf@atend #2 #3 .\\\else
   \ifx\epsf@tmp\epsf@atendlit\@atendtrue\fi\fi}

\chardef\psletter = 11 
\chardef\other = 12

\newif \ifdebug 
\newif\ifc@mpute 
\c@mputetrue 

\let\then = \relax
\def\r@dian{pt }
\let\r@dians = \r@dian
\let\dimensionless@nit = \r@dian
\let\dimensionless@nits = \dimensionless@nit
\def\internal@nit{sp }
\let\internal@nits = \internal@nit
\newif\ifstillc@nverging
\def \Mess@ge #1{\ifdebug \then \message {#1} \fi}

{ 
	\catcode `\@ = \psletter
	\gdef \nodimen {\expandafter \n@dimen \the \dimen}
	\gdef \term #1 #2 #3%
	       {\edef \t@ {\the #1}
		\edef \t@@ {\expandafter \n@dimen \the #2\r@dian}%
		\t@rm {\t@} {\t@@} {#3}%
	       }
	\gdef \t@rm #1 #2 #3%
	       {{%
		\count 0 = 0
		\dimen 0 = 1 \dimensionless@nit
		\dimen 2 = #2\relax
		\Mess@ge {Calculating term #1 of \nodimen 2}%
		\loop
		\ifnum	\count 0 < #1
		\then	\advance \count 0 by 1
			\Mess@ge {Iteration \the \count 0 \space}%
			\Multiply \dimen 0 by {\dimen 2}%
			\Mess@ge {After multiplication, term = \nodimen 0}%
			\Divide \dimen 0 by {\count 0}%
			\Mess@ge {After division, term = \nodimen 0}%
		\repeat
		\Mess@ge {Final value for term #1 of 
				\nodimen 2 \space is \nodimen 0}%
		\xdef \Term {#3 = \nodimen 0 \r@dians}%
		\aftergroup \Term
	       }}
	\catcode `\p = \other
	\catcode `\t = \other
	\gdef \n@dimen #1pt{#1} 
}

\def \Divide #1by #2{\divide #1 by #2} 

\def \Multiply #1by #2
       {{
	\count 0 = #1\relax
	\count 2 = #2\relax
	\count 4 = 65536
	\Mess@ge {Before scaling, count 0 = \the \count 0 \space and
			count 2 = \the \count 2}%
	\ifnum	\count 0 > 32767 
	\then	\divide \count 0 by 4
		\divide \count 4 by 4
	\else	\ifnum	\count 0 < -32767
		\then	\divide \count 0 by 4
			\divide \count 4 by 4
		\else
		\fi
	\fi
	\ifnum	\count 2 > 32767 
	\then	\divide \count 2 by 4
		\divide \count 4 by 4
	\else	\ifnum	\count 2 < -32767
		\then	\divide \count 2 by 4
			\divide \count 4 by 4
		\else
		\fi
	\fi
	\multiply \count 0 by \count 2
	\divide \count 0 by \count 4
	\xdef \product {#1 = \the \count 0 \internal@nits}%
	\aftergroup \product
       }}

\def\r@duce{\ifdim\dimen0 > 90\r@dian \then   
		\multiply\dimen0 by -1
		\advance\dimen0 by 180\r@dian
		\r@duce
	    \else \ifdim\dimen0 < -90\r@dian \then  
		\advance\dimen0 by 360\r@dian
		\r@duce
		\fi
	    \fi}

\def\Sine#1%
       {{%
	\dimen 0 = #1 \r@dian
	\r@duce
	\ifdim\dimen0 = -90\r@dian \then
	   \dimen4 = -1\r@dian
	   \c@mputefalse
	\fi
	\ifdim\dimen0 = 90\r@dian \then
	   \dimen4 = 1\r@dian
	   \c@mputefalse
	\fi
	\ifdim\dimen0 = 0\r@dian \then
	   \dimen4 = 0\r@dian
	   \c@mputefalse
	\fi
	\ifc@mpute \then
		\divide\dimen0 by 180
		\dimen0=3.141592654\dimen0
		\dimen 2 = 3.1415926535897963\r@dian 
		\divide\dimen 2 by 2 
		\Mess@ge {Sin: calculating Sin of \nodimen 0}%
		\count 0 = 1 
		\dimen 2 = 1 \r@dian 
		\dimen 4 = 0 \r@dian 
		\loop
			\ifnum	\dimen 2 = 0 
			\then	\stillc@nvergingfalse 
			\else	\stillc@nvergingtrue
			\fi
			\ifstillc@nverging 
			\then	\term {\count 0} {\dimen 0} {\dimen 2}%
				\advance \count 0 by 2
				\count 2 = \count 0
				\divide \count 2 by 2
				\ifodd	\count 2 
				\then	\advance \dimen 4 by \dimen 2
				\else	\advance \dimen 4 by -\dimen 2
				\fi
		\repeat
	\fi		
			\xdef \sine {\nodimen 4}%
       }}

\def\Cosine#1{\ifx\sine\UnDefined\edef\Savesine{\relax}\else
		             \edef\Savesine{\sine}\fi
	{\dimen0=#1\r@dian\advance\dimen0 by 90\r@dian
	 \Sine{\nodimen 0}
	 \xdef\cosine{\sine}
	 \xdef\sine{\Savesine}}}	      

\def\psdraft{
	\def\@psdraft{0}
}
\def\psfull{
	\def\@psdraft{100}
}

\psfull

\newif\if@scalefirst
\def\psscalefirst{\@scalefirsttrue}
\def\psrotatefirst{\@scalefirstfalse}
\psrotatefirst

\newif\if@draftbox
\def\psnodraftbox{
	\@draftboxfalse
}
\def\psdraftbox{
	\@draftboxtrue
}
\@draftboxtrue

\newif\if@prologfile
\newif\if@postlogfile
\def\pssilent{
	\@noisyfalse
}
\def\psnoisy{
	\@noisytrue
}
\psnoisy
\newif\if@bbllx
\newif\if@bblly
\newif\if@bburx
\newif\if@bbury
\newif\if@height
\newif\if@width
\newif\if@rheight
\newif\if@rwidth
\newif\if@angle
\newif\if@clip
\newif\if@verbose
\def\@p@@sclip#1{\@cliptrue}

\newif\if@decmpr

\def\@p@@sfigure#1{\def\@p@sfile{null}\def\@p@sbbfile{null}
	        \openin1=#1
		\ifeof1\closein1
	        	\openin1=\figurepath#1
			\ifeof1\closein1
			        \openin1=#1
				\ifeof1\closein1%
				       \openin1=\figurepath#1
					\ifeof1
					   \ps@typeout{Error, File #1 not found}
						\if@bbllx\if@bblly
				   		\if@bburx\if@bbury
			      				\def\@p@sfile{#1}%
			      				\def\@p@sbbfile{#1}%
							\@decmprfalse
				  	   	\fi\fi\fi\fi
					\else\closein1
				    		\def\@p@sfile{\figurepath#1}%
				    		\def\@p@sbbfile{\figurepath#1}%
						\@decmprfalse
	                       		\fi%
			 	\else\closein1%
					\def\@p@sfile{#1}
					\def\@p@sbbfile{#1}
					\@decmprfalse
			 	\fi
			\else
				\def\@p@sfile{\figurepath#1}
				\def\@p@sbbfile{\figurepath#1}
				\@decmprtrue
			\fi
		\else
			\def\@p@sfile{#1}
			\def\@p@sbbfile{#1}
			\@decmprtrue
		\fi}

\def\@p@@sfile#1{\@p@@sfigure{#1}}

\def\@p@@sbbllx#1{
		\@bbllxtrue
		\dimen100=#1
		\edef\@p@sbbllx{\number\dimen100}
}
\def\@p@@sbblly#1{
		\@bbllytrue
		\dimen100=#1
		\edef\@p@sbblly{\number\dimen100}
}
\def\@p@@sbburx#1{
		\@bburxtrue
		\dimen100=#1
		\edef\@p@sbburx{\number\dimen100}
}
\def\@p@@sbbury#1{
		\@bburytrue
		\dimen100=#1
		\edef\@p@sbbury{\number\dimen100}
}
\def\@p@@sheight#1{
		\@heighttrue
		\dimen100=#1
   		\edef\@p@sheight{\number\dimen100}
}
\def\@p@@swidth#1{
		\@widthtrue
		\dimen100=#1
		\edef\@p@swidth{\number\dimen100}
}
\def\@p@@srheight#1{
		\@rheighttrue
		\dimen100=#1
		\edef\@p@srheight{\number\dimen100}
}
\def\@p@@srwidth#1{
		\@rwidthtrue
		\dimen100=#1
		\edef\@p@srwidth{\number\dimen100}
}
\def\@p@@sangle#1{
		\@angletrue
		\edef\@p@sangle{#1} 
}
\def\@p@@ssilent#1{ 
		\@verbosefalse
}
\def\@p@@sprolog#1{\@prologfiletrue\def\@prologfileval{#1}}
\def\@p@@spostlog#1{\@postlogfiletrue\def\@postlogfileval{#1}}
\def\@cs@name#1{\csname #1\endcsname}
\def\@setparms#1=#2,{\@cs@name{@p@@s#1}{#2}}
\def\ps@init@parms{
		\@bbllxfalse \@bbllyfalse
		\@bburxfalse \@bburyfalse
		\@heightfalse \@widthfalse
		\@rheightfalse \@rwidthfalse
		\def\@p@sbbllx{}\def\@p@sbblly{}
		\def\@p@sbburx{}\def\@p@sbbury{}
		\def\@p@sheight{}\def\@p@swidth{}
		\def\@p@srheight{}\def\@p@srwidth{}
		\def\@p@sangle{0}
		\def\@p@sfile{} \def\@p@sbbfile{}
		\def\@p@scost{10}
		\def\@sc{}
		\@prologfilefalse
		\@postlogfilefalse
		\@clipfalse
		\if@noisy
			\@verbosetrue
		\else
			\@verbosefalse
		\fi
}
\def\parse@ps@parms#1{
	 	\@psdo\@psfiga:=#1\do
		   {\expandafter\@setparms\@psfiga,}}
\newif\ifno@bb
\def\bb@missing{
	\if@verbose{
		\ps@typeout{psfig: searching \@p@sbbfile \space  for bounding box}
	}\fi
	\no@bbtrue
	\epsf@getbb{\@p@sbbfile}
        \ifno@bb \else \bb@cull\epsf@llx\epsf@lly\epsf@urx\epsf@ury\fi
}	
\def\bb@cull#1#2#3#4{
	\dimen100=#1 bp\edef\@p@sbbllx{\number\dimen100}
	\dimen100=#2 bp\edef\@p@sbblly{\number\dimen100}
	\dimen100=#3 bp\edef\@p@sbburx{\number\dimen100}
	\dimen100=#4 bp\edef\@p@sbbury{\number\dimen100}
	\no@bbfalse
}
\newdimen\p@intvaluex
\newdimen\p@intvaluey
\def\rotate@#1#2{{\dimen0=#1 sp\dimen1=#2 sp
		  \global\p@intvaluex=\cosine\dimen0
		  \dimen3=\sine\dimen1
		  \global\advance\p@intvaluex by -\dimen3
		  \global\p@intvaluey=\sine\dimen0
		  \dimen3=\cosine\dimen1
		  \global\advance\p@intvaluey by \dimen3
		  }}
\def\compute@bb{
		\no@bbfalse
		\if@bbllx \else \no@bbtrue \fi
		\if@bblly \else \no@bbtrue \fi
		\if@bburx \else \no@bbtrue \fi
		\if@bbury \else \no@bbtrue \fi
		\ifno@bb \bb@missing \fi
		\ifno@bb \ps@typeout{FATAL ERROR: no bb supplied or found}
			\no-bb-error
		\fi
		\count203=\@p@sbburx
		\count204=\@p@sbbury
		\advance\count203 by -\@p@sbbllx
		\advance\count204 by -\@p@sbblly
		\edef\ps@bbw{\number\count203}
		\edef\ps@bbh{\number\count204}
		\if@angle 
			\Sine{\@p@sangle}\Cosine{\@p@sangle}
	        	{\dimen100=\maxdimen\xdef\r@p@sbbllx{\number\dimen100}
					    \xdef\r@p@sbblly{\number\dimen100}
			                    \xdef\r@p@sbburx{-\number\dimen100}
					    \xdef\r@p@sbbury{-\number\dimen100}}
                        \def\minmaxtest{
			   \ifnum\number\p@intvaluex<\r@p@sbbllx
			      \xdef\r@p@sbbllx{\number\p@intvaluex}\fi
			   \ifnum\number\p@intvaluex>\r@p@sbburx
			      \xdef\r@p@sbburx{\number\p@intvaluex}\fi
			   \ifnum\number\p@intvaluey<\r@p@sbblly
			      \xdef\r@p@sbblly{\number\p@intvaluey}\fi
			   \ifnum\number\p@intvaluey>\r@p@sbbury
			      \xdef\r@p@sbbury{\number\p@intvaluey}\fi
			   }
			\rotate@{\@p@sbbllx}{\@p@sbblly}
			\minmaxtest
			\rotate@{\@p@sbbllx}{\@p@sbbury}
			\minmaxtest
			\rotate@{\@p@sbburx}{\@p@sbblly}
			\minmaxtest
			\rotate@{\@p@sbburx}{\@p@sbbury}
			\minmaxtest
			\edef\@p@sbbllx{\r@p@sbbllx}\edef\@p@sbblly{\r@p@sbblly}
			\edef\@p@sbburx{\r@p@sbburx}\edef\@p@sbbury{\r@p@sbbury}
		\fi
		\count203=\@p@sbburx
		\count204=\@p@sbbury
		\advance\count203 by -\@p@sbbllx
		\advance\count204 by -\@p@sbblly
		\edef\@bbw{\number\count203}
		\edef\@bbh{\number\count204}
}
\def\in@hundreds#1#2#3{\count240=#2 \count241=#3
		     \count100=\count240	
		     \divide\count100 by \count241
		     \count101=\count100
		     \multiply\count101 by \count241
		     \advance\count240 by -\count101
		     \multiply\count240 by 10
		     \count101=\count240	
		     \divide\count101 by \count241
		     \count102=\count101
		     \multiply\count102 by \count241
		     \advance\count240 by -\count102
		     \multiply\count240 by 10
		     \count102=\count240	
		     \divide\count102 by \count241
		     \count200=#1\count205=0
		     \count201=\count200
			\multiply\count201 by \count100
		 	\advance\count205 by \count201
		     \count201=\count200
			\divide\count201 by 10
			\multiply\count201 by \count101
			\advance\count205 by \count201
		     \count201=\count200
			\divide\count201 by 100
			\multiply\count201 by \count102
			\advance\count205 by \count201
		     \edef\@result{\number\count205}
}
\def\compute@wfromh{
		\in@hundreds{\@p@sheight}{\@bbw}{\@bbh}
		\edef\@p@swidth{\@result}
}
\def\compute@hfromw{
	        \in@hundreds{\@p@swidth}{\@bbh}{\@bbw}
		\edef\@p@sheight{\@result}
}
\def\compute@handw{
		\if@height 
			\if@width
			\else
				\compute@wfromh
			\fi
		\else 
			\if@width
				\compute@hfromw
			\else
				\edef\@p@sheight{\@bbh}
				\edef\@p@swidth{\@bbw}
			\fi
		\fi
}
\def\compute@resv{
		\if@rheight \else \edef\@p@srheight{\@p@sheight} \fi
		\if@rwidth \else \edef\@p@srwidth{\@p@swidth} \fi
}
\def\compute@sizes{
	\compute@bb
	\if@scalefirst\if@angle
	\if@width
	   \in@hundreds{\@p@swidth}{\@bbw}{\ps@bbw}
	   \edef\@p@swidth{\@result}
	\fi
	\if@height
	   \in@hundreds{\@p@sheight}{\@bbh}{\ps@bbh}
	   \edef\@p@sheight{\@result}
	\fi
	\fi\fi
	\compute@handw
	\compute@resv}

\def\psfig#1{\vbox {
	\ps@init@parms
	\parse@ps@parms{#1}
	\compute@sizes
	\ifnum\@p@scost<\@psdraft{
		\special{ps::[begin] 	\@p@swidth \space \@p@sheight \space
				\@p@sbbllx \space \@p@sbblly \space
				\@p@sbburx \space \@p@sbbury \space
				startTexFig \space }
		\if@angle
			\special {ps:: \@p@sangle \space rotate \space} 
		\fi
		\if@clip{
			\if@verbose{
				\ps@typeout{(clip)}
			}\fi
			\special{ps:: doclip \space }
		}\fi
		\if@prologfile
		    \special{ps: plotfile \@prologfileval \space } \fi
		\if@decmpr{
			\if@verbose{
				\ps@typeout{psfig: including \@p@sfile \space }
			}\fi
			\special{ps: plotfile \@p@sfile \space }
		}\else{
			\if@verbose{
				\ps@typeout{psfig: including \@p@sfile \space }
			}\fi
			\special{ps: plotfile \@p@sfile \space }
		}\fi
		\if@postlogfile
		    \special{ps: plotfile \@postlogfileval \space } \fi
		\special{ps::[end] endTexFig \space }
		\vbox to \@p@srheight sp{
			\hbox to \@p@srwidth sp{
				\hss
			}
		\vss
		}
	}\else{
		\if@draftbox{		
			\hbox{\frame{\vbox to \@p@srheight sp{
			\vss
			\hbox to \@p@srwidth sp{ \hss \@p@sfile \hss }
			\vss
			}}}
		}\else{
			\vbox to \@p@srheight sp{
			\vss
			\hbox to \@p@srwidth sp{\hss}
			\vss
			}
		}\fi

	}\fi
}}
\psfigRestoreAt
\let\@=\LaTeXAtSign

\tightenlines
\begin{document}

\twocolumn[
\hsize\textwidth\columnwidth\hsize\csname@twocolumnfalse\endcsname
%
%

%
\draft
 
\title{Resonant Inelastic X-Ray Scattering from Valence Excitations in 
insulating Copper-Oxides}
 
\author{P.~Abbamonte$^{1,2}$, C.~A.~Burns$^3$, E.~D.~Isaacs$^2$,
P.~M.~Platzman$^2$, L.~L.~Miller$^4$, 
S.~W.~Cheong$^2$, and M.~V.~Klein$^1$\\}
 
\address{
$^1$Department of Physics, University of Illinois, 1110 W. Green St., Urbana, IL, 61801\\
$^2$Bell Laboratories, Lucent Technologies, 600 Mountain Av., Murray Hill, NJ, 07974\\
$^3$Department of Physics, Western Michigan University, Kalamazoo, MI, 49008\\
$^4$Ames Laboratory, Ames, IA, 50011\\
}
 
\date{\today}
\maketitle
 
\begin{abstract}
We report resonant inelastic x-ray measurements of 
insulating La$_2$CuO$_4$ and 
Sr$_2$CuO$_2$Cl$_2$ taken with the incident energy tuned near the
Cu K absorption edge.  We show that the spectra are well
described in a shakeup picture in 3rd order perturbation theory which
exhibits both incoming and outgoing resonances, and demonstrate how to 
extract a spectral function from the raw data.
We conclude by showing {\bf q}-dependent measurements of the charge
transfer gap.
\end{abstract}
 
]
\narrowtext
Inelastic x-ray scattering (IXS) has shown promise as a practical probe of 
electronic excitations in condensed matter because of its broad 
kinematic range and direct
coupling to the electron charge.  However, since x-rays are strongly
absorbed in high density materials, successful applications of the technique
have been limited to low-Z 
systems\cite{krisch,johnhillli,schulke1,benlarson,v2o3}.

Several recent studies\cite{eric,chichang,johnhill,butorin},
have shown that by that by tuning the incident energy near an x-ray
absorption edge a Raman effect could be measured, despite the high 
absorption, because of the resonant enhancement.  These studies have
emphasized the role of coulomb interactions in the scattering 
process.  Since it involves coupling between highly excited
virtual states and strongly correlated valence states, it is
important to characterize the resonance process well for the
technique to be useful.

With emphasis on systematics, we have measured resonant inelastic x-ray
scattering at moderate resolution ($\Delta E$=0.9 eV) near the CuK
absorption edge in the high-T$_c$ parent insulator La$_2$CuO$_4$ (LCO) as
a function of incident photon energy.  Based on the changes of inelastic
intensity and peak position with incident energy we show that the scattering
is well described in a ``shakeup" picture in 3rd order perturbation
threoy\cite{phil}.  We also present higher resolution measurements
($\Delta E$=0.45 eV) on another insulator, Sr$_2$CuO$_2$Cl$_2$ 
(SCOC), as a function of momentum transfer, {\bf q}, which show
some new features, such as the 2 eV optical charge transfer gap.

Experiments were carried out at the X21 wiggler line at the National
Synchrotron Light Source and the 3ID (SRI-CAT) undulator line at the 
Advanced Photon Source.  At X21 the energy resolution was 0.45 eV and
typical count rates were 0.4 Hz.  At 3ID with 0.9 eV resolution 9 Hz was
typical.  The scattered light was collected with a spherical, diced, 
Ge(733) analyzer and imaged onto a detector.  Energy analysis was done
by rotating the analyzer and translating the detector in coincidence.
The momentum transfer was varied by rotating the entire apparatus around
the scattering center (exact experimental geometries are indicated
in the figure captions).

\begin{figure} 
\begin{center}
  \mbox{\psfig{figure=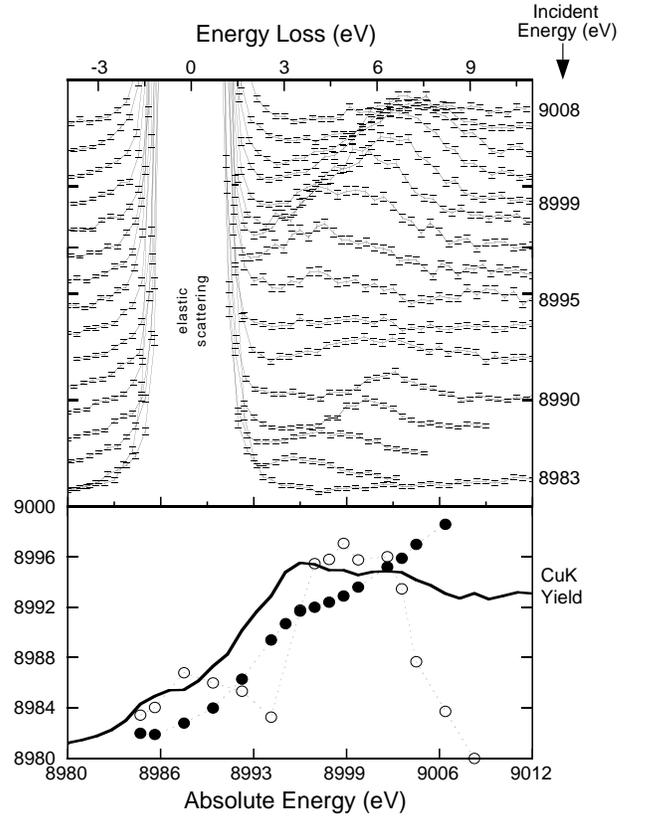,width=3.2in,silent=}}
   \caption{
Resonance Profile for La$_2$CuO$_4$ taken 
at 3ID with $\hat{\epsilon}_i||\hat{a}$ and
q=1.27$\AA^{-1}$ parallel to $\hat{c}$.  The upper 
frame shows the raw spectra plotted against {\it transferred} 
energy (curves are offset for clarity).
In the lower frame the open and filled circles are the 
inelastic peak height and
position, respectively, plotted against {\it incident} energy.  
The black line is the
fluorescence yield which peaks at the Cu$1s\rightarrow 4p$ energy.
}
\end{center}
\end{figure}

The LCO and SCOC crystals were grown by techniques described previously
\cite{lance,cheong}.  They were characterized with a spectroscopic
ellipsometer to assure surface quality.

In the upper panel of Fig. 1 we show inelastic x-ray 
spectra from La$_2$CuO$_4$ at fixed {\bf q} for 16 different 
incident x-ray energies across the Cu K absorption edge. 
Since the overall absorption change across the 
edge is only 14\% (x-ray absorption being dominated by the La 
atoms) corrections for sample self-absorption did not
significantly alter the line shapes.  So shown here are the raw spectra.
The strong peak at zero energy
loss is elastic scattering.  The energy gain side (negative energy loss) 
shows a background
of 12 counts per minute.  
The principal feature of each 
curve is a single peak, seen previously in Nd$_2$CuO$_4$\cite{johnhill},
whose position, intensity, and line shape vary greatly as a function of 
incident energy.  

We summarize the resonant behavior 
in the lower frame of Fig. 1 where the inelastic peak height 
(open circles) and position (filled circles) are plotted against
{\it incident} energy.  The thick line is the CuK fluorescence
yield which shows the location of the edge (a localized
Cu$1s\rightarrow 4p$ transition).  The peak height shows two maxima, 
the stronger of which correlates with 
the peak of the edge and the weaker 
with the pre-edge shoulder.  In both cases the maximum 
is offset from the absorption peak by about 2.5 eV.

The peak {\it position} shifts nonlinearly with incident energy;
below and above a resonance it is roughly linear while near a
resonance it plateaus.  This behavior differs fundamentally
from a classic Raman effect, in which one expects a linear peak shift with
unit slope with a gradual rise in intensity below a resonance, and 
saturation above \cite{eisenberger}.

Independent of any model, this type of low energy loss scattering leads
to excited states of the valence electron system in the absence of
core excitations.  However, because one is near the $1s$ absorption 
threshold of Cu the scattering proceeds through a set of almost real,
highly excited (9 keV) intermediate states, which have a $1s$ core hole
and an extra electron excited in a localized $4p$ state ($\bar{1s}4p$).  
When the highly excited intermediate state disappears it leaves behind
low-lying valence excited states - in principle conserving energy and crystal
momentum.

One must describe these many-body intermediate states, 
i.e. their matrix elements
as well as their off-shell weight, in order to characterize the coherent,
second order process.  Different groups have resorted to different 
approximation schemes.  Starting with N valence electrons in a small cluster,
Tanaka and Kotani \cite{kotaniCeO} describe the intermediate
states as a set of N+1 interacting electrons
in the presence of a rigid impurity - the
Anderson Impurity Model.  
This treatment assumes that the core state is suddenly created, and that
it can be treated as a fixed, external potential.  It
emphasizes the multiplet coupling among the N+1 electrons, and
is done numerically in exact diagonalization with a large number 
of basis functions.

Taking a more analytic approach, 
Platzman and Isaacs\cite{phil} treat the many-body
problem by describing Coulomb interactions among electrons in a series of
perturbation diagrams\cite{gelmukhanov}.
This approach assumes that interactions can be taken to be weak for
a suitable choice of basis functions {\it or}
that one can sum enough terms in perturbation theory to include 
the important physics.  They argue that near a sharp, dipole-allowed
transition the dominant term occurs in 3rd
order.  Writing it out explicitly one arrives at

\begin{equation}
S_{f\leftarrow i}=\sum_{\bar{1s},4p} \frac{M_{em} \; M_{coul} \; M_{abs}}
{(\omega_s - E_{\bar{1s},4p} + i\gamma_{K})(\omega_i-E_{\bar{1s},4p}+i\gamma_K)}
\end{equation}

\noindent
In this expression the sum is on all possible states of the $1s$ hole and
$4p$ electron, $E_{\bar{1s}4p}$ is their energy, and $\gamma_K$ is their inverse 
lifetime.  The numerator contains matrix elements for absorption, 
M$_{abs}=(e/mc)\langle\bar{1s}\,4p|{\bf p}\cdot{\bf \hat{A}}|{\bf k}_i\rangle$,
emission, M$_{em}=(e/mc)\langle {\bf k}_s|{\bf p}\cdot{\bf \hat{A}}
|\bar{1s}' \, 4p'\rangle$,
and coulomb interaction between the core and valence states,
M$_{coul}=\int{d{\bf x}\,d{\bf x'} \, \langle\bar{1s}' \, 4p';f|\hat{\rho}({\bf x})
\hat{\rho}({\bf x}')/|{\bf x}-{\bf x'}||\bar{1s} \, 4p;i\rangle}$.
{\bf k}$_i$ and {\bf k}$_s$ are the incident and scattered photon momenta,
(i.e. {\bf q}={\bf k}$_i$-{\bf k}$_s$) and $\omega=\omega_i-\omega_s$ is the energy loss.

Physically this expression represents the following.  The incident photon,
with energy tuned to the CuK absorption edge, creates a virtual $\bar{1s}4p$
pair on a Cu site.  This pair is bound as an exciton by the coulomb 
interaction (not included in Eq. (1)) and so is non-dispersive.  It takes up the
momentum of the incident photon, 4.55 $\AA^{-1}$, and scatters off the
valence electron system.  When the exciton recombines, the emitted photon
reflects the energy and momentum imparted to the system.  This is commonly
called a ``shakeup" process, which to first order in the coulomb interaction
is given by Eq. (1).  

To get a transition rate we square the quantity (1) and perform an
incoherent sum on final states.  We postulate that the intermediate states
are approximately degenerate with energy $E_{\bar{1s}4p}=E_K$
(since they are spacially localized),
which allows factoring of the energy denominators from the sum.
We arrive at

\begin{equation}
w=\frac{S_K({\bf q},\omega;\hat{\epsilon}_i,\hat{\epsilon}_s)}
{\left [ (\omega_i-E_K)^2+{\gamma_K}^2 \right ] \, 
\left [ (\omega_s-E_K)^2+{\gamma_K}^2 \right]}
\end{equation}

\noindent
where

\begin{equation}
S_K=\frac{2\pi}{\hbar}\sum_{f}
\left |\sum_{\bar{1s},4p}\,M_{em}\,M_{abs}\,M_{coul}\right |^2
\delta(\omega-E_f+E_i).
\end{equation}

\noindent
and $\hat{\bf \epsilon}_i$ and $\hat{\bf \epsilon}_s$
are the polarizations of the
incident and scattered photons.  The two lorentzians in (2) are
incoming and outgoing resonances in the photon frequency, so we
see that this treatment is completely analogous to third-order
optical Raman scattering from phonons in semiconductors,
in which the scattering is described 
by a single operation of the {\it electron-phonon}
interaction on a virtual {\it valence} electron-hole pair\cite{cardona}.
The two resonances formally come about the same way.
 
The central result of this paper is Eq. (2).  It says that,
within our approximation, all the very different 
spectra in Figure 1 derive from the same fundamental 
quantity, $S_K$, which depends on the {\it difference}
$\omega=\omega_i-\omega_s$ rather than on $\omega_i$ and $\omega_s$
independently.  A way to test this result
would be to take the curves from Fig. 1, divide each by its
respective denominator from Eq. (2), and see if they all
collapse to the same function.

\begin{figure}
\begin{center}
\mbox{\psfig{figure=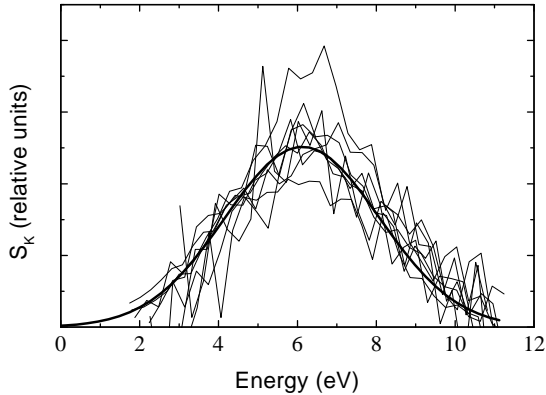,width=3.2in,silent=}}
\caption{
The result of dividing energy denominators in (2) from the experimental 
spectra in Fig. 1.  
For $E_K=(8995.14\pm 1.61)$ eV and $\gamma_K=(2.38\pm 0.542)$ eV
all spectra collapse onto a single curve.  The thick line is a Gaussian fit.
}
\end{center}
\end{figure}

In Eq. (2) we assumed that we are near a single resonance, so
we take the nine highest curves from Fig. 1 (around the second peak 
in the resonance profile) and subtract their background and elastic
scattering.  
We use $E_K$ and $\gamma_K$ as flexible parameters and divide
each spectrum by its respective denominator.  Using a nonlinear
fitting algorithm, we adjust the values of $E_K$ and $\gamma_K$
to minimize the total variation (the $\chi^2$ summed over all
points and all spectra) among the curves, irrespective of the
resulting shape.
For the values $E_K=(8995.14\pm 1.61)$ eV and 
$\gamma_K=(2.38\pm 0.542)$ eV we find that the
spectra collapse onto a single curve, shown in Figure 2.  

The result for $S_K({\bf q},\omega;\hat{\epsilon}_i,\hat{\epsilon}_s)$ is
a single peak at 6.1 eV energy loss and width of 3.9 eV.  Referring to
the cluster calculations of S\'{\i}mon \cite{simon} we suggest that this
feature is a transition from the $b_{1g}$ ground state to an $a_{1g}$
excitonic excited state composed of symmetric combinations of a central
Cu3$d_{x^2-y^2}$ orbital and the surrounding O2$p_{\sigma}$ orbitals.

To illustrate what {\it qualitative} aspects of the data are captured by our fit,
i.e. by the resonant denominators in (2), we take
a single function for $S_K$, i.e. a fit to the collapsed data in Fig. 2 
(thick line), combine it with the denominators in Eq. (2), 
and produce the model resonance profile shown in Fig. 3
(identical formatting to Fig. 1).
The salient features are reproduced, including the
peak shift with incident energy and
the 2.5 eV offset. 
All this behavior comes from the energy denominators in (2) and is
independent of the nature of the core state or the 
particular valence excitation.

\begin{figure} 
\begin{center}
  \mbox{\psfig{figure=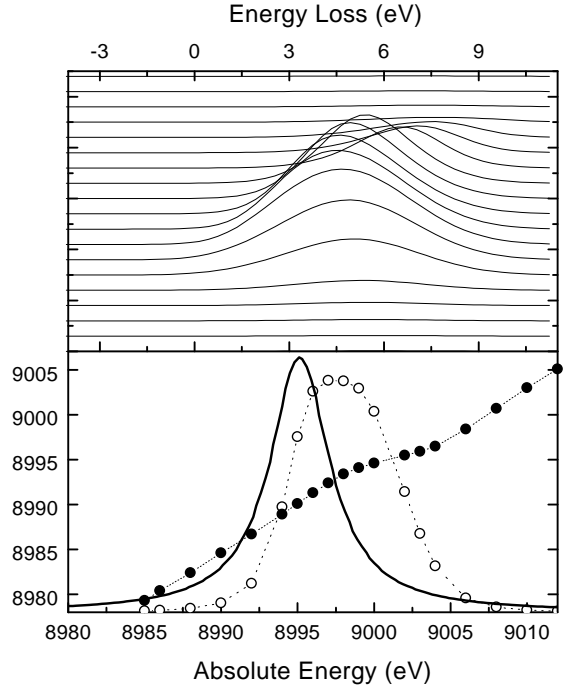,width=3in,silent=}}
   \caption{
A model resonance profile generated from the dark line in
Fig. 2 and the two resonant denominators in equation (2). Formatting is
identical to Fig. 1.
The solid line is a simulated CuK fluorescence yield showing the absorption 
peak.  The continuum above the peak is omitted since it 
does not contribute to the resonance.
}
\end{center}
\end{figure}

In this simple shakeup description 
$S_K({\bf q},\omega;\hat{\epsilon}_i,\hat{\epsilon}_s)$ has an
explicit relationship with $S({\bf q},\omega)$, 
the dynamical structure factor measured 
in nonresonant inelastic
x-ray scattering\cite{schulke2}.  
This can be seen by writing out the matrix element
$M_{coul}$ in momentum space, which (neglecting exchange between
core and valence states) has the form

\begin{equation}
M_{coul}=\sum_{\bf G} \frac{4\pi e^2}{|{\bf q}+{\bf G}|^2}
F_{\bar{1s}4p}({\bf q}+{\bf G};\hat{\epsilon_i})
\langle f|\hat{\rho}_{v,{\bf q}+{\bf G}}|i\rangle.
\end{equation}

\noindent
Here $F_{\bar{1s}4p}(k)$ is the static x-ray form factor of the
$\bar{1s}4p$ exciton.  It is dependent implicitly on the incident
polarization $\hat{\epsilon_i}$ since in the dipole approximation
$M_{abs}$ determines the spacial orientation of the $4p$. 
$\hat{\rho}_{v,{\bf q}}$
is the valence part of the many body density operator 
$\hat{\rho}_{\bf q}$ and 
the sum in (4) is on all reciprocal lattice vectors, {\bf G}.

The quantity $\langle f|\hat{\rho}_{v,{\bf q}}|i\rangle$,
when squared, multiplied by $\delta(\omega-E_f+E_i)$, 
and summed on final states, 
is identically the valence part of 
the dynamic structure factor $S({\bf q},\omega)$.  
Doing the sum on {\bf G} before squaring we find that 
$S_K({\bf q},\omega;\hat{\epsilon}_i,\hat{\epsilon}_s)$ 
is a superposition of many $S({\bf q}+{\bf G},\omega)$ 
functions, weighted by the form 
factor of the core states.  Therefore, $S_K$ can be thought of as
a response function similar to $S$ ``projected" onto the form factor
of the core state. 

\begin{figure}
\begin{center}
\mbox{\psfig{figure=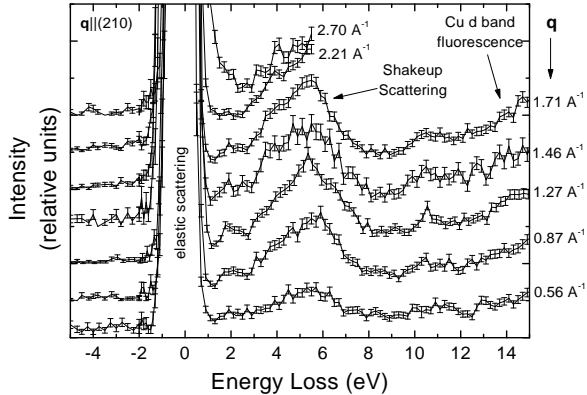,width=8.5cm,silent=}}
\caption{
Resonant spectra from
Sr$_2$CuO$_2$Cl$_2$ as a function of {\bf q}, taken at X21 with
both {\bf q} and {\bf E} approximately parallel to the (210)
crystallographic direction.
The feature at 2 eV is the charge transfer gap, which changes shape and
shifts with increasing ${\bf q}$.
}
\end{center}
\end{figure}

Finally we present some higher-resolution ($\Delta$E=0.45 eV),
{\bf q}-dependent spectra
from the insulator Sr$_2$CuO$_2$Cl$_2$ (Fig. 4).
The salient spectral features are three peaks: the same feature we
saw in LCO (appearing here at about 5 eV), a feature at 2 eV
which appears to shift and lose definition with {\bf q}, and a peak
at 10.5 eV which is absent at low-{\bf q} but which gains intensity
as {\bf q} is increased.  The 2 eV feature strongly resembles
dipole-active charge transfer gap seen with optical 
reflectivity\cite{zibold}.

To summarize, we measured RIXS spectra at the
CuK edge in LCO and SCOC.  From the resonance profile we deduce
that the scattering can be described as a shakeup process in 3rd
order which, analogous to optical Raman scattering from 
phonons, exhibits incoming and outgoing
resonances.  The scattered intensity has the form of
a doubly resonant form factor multiplied by a response function, 
$S_K({\bf q},\omega;\hat{\bf \epsilon}_i,\hat{\bf \epsilon}_s)$, 
which can be thought of as the dynamical structure factor
projected onto the form factor of the intermediate core state.

The advantages of this description are (i) that it makes no assumption that the
core state is rigid and so momentum conservation enters naturally, and (ii) 
that it allows one, given certain knowledge of the intermediate state 
(i.e. $F_{\bar{1s}4p}({\bf k})$), to relate the scattering to a response 
function {\it in terms of the valence electrons only}.  This 
description is useful when the core resonance is sharp 
and well isolated, and when one is mostly 
interested in the valence electron spectrum and is
willing to sacrifice a detailed multiplet description 
of the core states.

We gratefully acknowledge E. E. Alp, Z. Hasan, C.-C. Kao, 
V. I. Kushnir, P. L. Lee, H. L. Liu, A. T. Macrander, G. A. Sawatzky,
M. Schwoerer-B\"{o}hning, M. E. S\'{\i}mon,
S. K. Sinha, J. P. Sutter, T. Toellner, and C. Varma.
This work was supported by the NSF under grant DMR-9705131 and 
by the U.S. Department
of Energy, Basic Energy Sciences, Office of Energy Research, under
contract no. W-31-109-ENG-38.

\end{document}